\documentclass[12pt,usenames,dvipsnames,a4paper]{article}

\usepackage{soul}

\usepackage[utf8]{inputenc}
\usepackage[T1]{fontenc}

\usepackage{changepage}
\usepackage{amsmath,amsfonts,amssymb}
\usepackage{epsf,bbold,stmaryrd}
\usepackage{mathrsfs}
\usepackage{appendix}
\usepackage{caption}
\usepackage{color}
\usepackage{datetime}
\usepackage{float}
\usepackage{graphicx}
\usepackage[colorlinks,hyperindex,breaklinks]{hyperref}
\hypersetup{pageanchor=false,citecolor=red,urlcolor=red}
\usepackage{indentfirst}
\usepackage[numbers,square,comma,sort&compress,merge]{natbib} 
\usepackage{subcaption}
\usepackage{mathtools}
\usepackage{ytableau}
\usepackage{tabu}
\usepackage{esvect}

\usepackage{calc}
\usepackage{bm}

\allowdisplaybreaks

\def\a{\alpha}

\def\b{\beta}
\def\c{\chi}

\def\eps{\varepsilon}
\def\f{\frac}

\def\l{\left}

\def\mc{\mathcal}

\def\m{\mu}
\def\n{\nu}

\def\nn{\nonumber}

\def\Om{\Omega}

\def\p{\partial}

\def\r{\right}
\def\s{\sigma}

\def\x{\xi}

\def\be{\begin{equation}}
\def\ee{\end{equation}}

\def\bea{\begin{eqnarray}}
\def\eea{\end{eqnarray}}

\def\ba{\begin{array}}
\def\ea{\end{array}}

\def\bc{\begin{center}}
\def\ec{\end{center}}

\def\bl{\begin{flushleft}}
\def\el{\end{flushleft}}

\def\br{\begin{flushright}}
\def\er{\end{flushright}}

\def\bi{\begin{itemize}}
\def\ei{\end{itemize}}

\def\bt{\begin{tabular}}
\def\et{\end{tabular}}

\makeatletter

\newsavebox\myboxA
\newsavebox\myboxB
\newlength\mylenA

\newcommand*\xoverline[2][0.75]{%
    \sbox{\myboxA}{$\m@th#2$}%
    \setbox\myboxB\null
    \ht\myboxB=\ht\myboxA%
    \dp\myboxB=\dp\myboxA%
    \wd\myboxB=#1\wd\myboxA
    \sbox\myboxB{$\m@th\overline{\copy\myboxB}$}
    \setlength\mylenA{\the\wd\myboxA}
    \addtolength\mylenA{-\the\wd\myboxB}%
    \ifdim\wd\myboxB<\wd\myboxA%
       \rlap{\hskip 0.5\mylenA\usebox\myboxB}{\usebox\myboxA}%
    \else
        \hskip -0.5\mylenA\rlap{\usebox\myboxA}
         {\hskip 0.5\mylenA\usebox\myboxB}%
    \fi}
\makeatother

\usepackage{xspace}
\newcommand*{\ie}{i.e., }
\newcommand*{\eg}{e.g., }

\usepackage{xifthen}
\newcommand*\diff{\mathrm{d}} 
\newcommand*\ldiff[2][]{ \ifthenelse{\isempty{#1}}{ \diff
#2}{\diff^#1#2} \,} 
\let\limitint\int 
\renewcommand{\int}{\limitint \!} 

\begin{document}

\begin{titlepage}
\vspace{5cm}

\vspace{2cm}

\begin{center}
    \Large\textbf{Three-form lifting of dilaton flat direction \\
    without and with gravity}
\end{center}

\vspace{1cm}

\begin{center}
{\textsc {Georgios K. Karananas}}
\end{center}

\begin{center}
{\it Max-Planck-Institut für Physik\\ 
Boltzmannstraße 8, 85748 Garching bei M\"unchen, Germany\\
\vspace{.2cm}
Arnold Sommerfeld Center\\
Ludwig-Maximilians-Universit\"at M\"unchen\\
Theresienstra{\ss}e 37, 80333 M\"unchen, Germany}\\

\end{center}

\begin{center}
\texttt{\small georgios.karananas@physik.uni-muenchen.de} \\
\end{center}

\vspace{2cm}
\begin{abstract}

Spontaneous scale symmetry breaking is commonly associated with a flat
direction in the action. We show that this need not be so if the dilaton is
coupled to a three-form field in a manner compatible with gauge invariance and
dilatations. The resulting effective dynamics lifts the flat direction without
introducing explicit scale-violating operators. When gravity is included, the
corresponding potential takes the form of an exponential plateau.

\end{abstract}

\end{titlepage}

\section{Introduction}

The spontaneous breakdown of exact scale symmetry in a Poincar\'e-invariant
fashion translates into a flat direction in the effective potential. For a
single real scalar field this means that its quartic self-coupling must
vanish. The purpose of this paper is to show that the presence of a three-form
gauge sector bypasses this conclusion in a simple and fully scale-invariant
way.

In four spacetime dimensions, a three-form carries no propagating degrees of
freedom. Its equation of motion, however, introduces a dimensionful
integration constant. When the dilaton is coupled to such a gauge sector in a
manner compatible with dilatations---this differs crucially from the usual
shift-symmetric interaction familiar from axion
physics~\cite{Dvali:2005an}\footnote{This mechanism has led to important
insights on the axion solution to the strong CP
problem~\cite{Dvali:2005an,Dvali:2017mpy,Dvali:2022fdv}, and even found
applications in natural
inflation~\cite{Kaloper:2008fb,Kaloper:2011jz}.}---this constant feeds into
the scalar potential that develops a nontrivial minimum, and the flat
direction is lifted. No explicit scale-breaking operator is needed.
Dilatations remain exact at the level of the action; the scale is selected
dynamically by the ``branch'' picked out by the three-form equation of motion.

This mechanism is already operative in flat spacetime, where the three-form
generates a symmetry-breaking quartic potential for the dilaton, singling out
a nonzero vacuum expectation value. Once gravity is included, however, the
same considerations are altered in an interesting way. Scale symmetry forces
the dilaton to couple nonminimally to the Ricci scalar, and the three-form
induced potential is reshaped into an exponentially flat plateau, placing such
a construction in the same universality class as Higgs and Starobinsky
inflation. This contrasts sharply with scale-invariant unimodular gravity,
where the analogous integration constant induces a potential for the dilaton
of the runaway type. Hence, the field assumes the role of
quintessence~\cite{Shaposhnikov:2008xb,Garcia-Bellido:2011kqb}, very much in
the spirit of the cosmon
scenario~\cite{Peccei:1987mm,Wetterich:1987fm,Wetterich:1994bg}.
Interestingly, in the presence of exact scale symmetry, the usual
correspondence between the unimodular and three-form descriptions no longer
holds.

This paper is organized as follows. In Sec.~\ref{sec:flat}, we review some
basics of three-forms in flat spacetime and show how a scale-invariant
coupling to the dilaton lifts the flat direction and triggers spontaneous
breaking of dilatations. In Sec.~\ref{sec:gravity}, we include gravity, and
demonstrate how the potential becomes exponentially flat. We then compare the
construction with scale-invariant unimodular gravity, and also briefly comment
on a purely gravitational realization of the same idea in pure $R^2$ gravity,
where the Einstein--Hilbert term can be dynamically generated. We conclude in
Sec.~\ref{sec:conclusions}.

We work in four spacetime dimensions. The metric is taken to be mostly plus.
For the Levi-Civita symbol we use $\eps^{0123} = -1$.

\section{Flat spacetime}
\label{sec:flat}

\subsection{Three-form basics}

We begin by recapping some basic facts about three-forms, see
e.g.~\cite{Dvali:1995ce,Kaloper:2008fb,Kaloper:2011jz}, and fix notation and
conventions.

Consider a three-form gauge field $C_{\m\n\rho}$; its field strength is given
by
\be
\label{eq:field_strength}
F_{\m\n\rho\s}=4\,\p_{[\m}C_{\n\rho\s]} \ , 
\ee
where the square brackets stand for normalized antisymmetrization of the
enclosed indexes. 

The above is manifestly invariant under the gauge redundancy 
\be
\label{eq:C_gauge_symmetry}
C_{\m\n\rho} \to  C_{\m\n\rho} + 3\p_{[\m} \Om_{\n\rho]} \ ,
\ee
with $\Om_{\m\n} = - \Om_{\n\m}$. This ensures that the three-form carries no
propagating degrees of freedom (in four dimensions). 

Consider the lowest-order gauge invariant action involving the three-form in
flat spacetime, which reads
\be
\label{eq:3form_flat_action}
S_{\rm 3-form} =  
\int \diff^4x\l[- \f{1}{48} F_{\m\n\rho\s}F^{\m\n\rho\s}  
+ \f 1 6 \p_\m\l(F^{\m\n\rho\s}C_{\n\rho\s}\r)\r] \ ,
\ee
and we explicitly included a total derivative. Although this does not
contribute to the equations of motion, it is nonetheless essential for a
well-posed variational principle when the field strength is to be held fixed
on the boundary~\cite{Duff:1989ah,Duncan:1989ug}. 

Alongside the three-form gauge redundancy, the
action~(\ref{eq:3form_flat_action}) also enjoys invariance under global scale
transformations\,\footnote{The three-form theory, however, is not invariant
under special conformal transformations unless the spacetime is
eight-dimensional. This can be seen, \eg by explicitly computing the conformal
variation of~(\ref{eq:3form_flat_action}); or by showing that the
energy-momentum tensor cannot be improved, since the virial current cannot be
expressed as a total derivative; or, what amounts to the same thing, by
demonstrating that the theory cannot be coupled to gravity in a Weyl invariant
manner (see, however, Ref.~\cite{Karananas:2015ioa} for exceptions to this
rule of thumb). Thus, it provides another example of a unitary theory which is
scale invariant without being conformal, much like Maxwell theory in $d\neq
4$~\cite{El-Showk:2011xbs}. }
\be
\label{eq:dilatations_CF}
C_{\m\n\rho}(x) 
\to  C'_{\m\n\rho}(x) =  e^{\s}C_{\m\n\rho}(e^\s x) \ ,
~~~F_{\m\n\rho\s}(x) 
\to  F'_{\m\n\rho\s}(x) = e^{2\s}F_{\m\n\rho\s}(e^\s x) \ ,
\ee
with $\s$ a real parameter.  

To make transparent how dilatations get broken in this construction, it is
more convenient to work with the first order
formulation~\cite{Kaloper:2008fb,Kaloper:2011jz} and treat $F_{\m\n\rho\s}$
and $C_{\m\n\rho}$ as independent. We thus consider the following equivalent
form of the action~(\ref{eq:3form_flat_action}) 
\be
\label{eq:3form_first_order}
S_{\rm 3-form} = 
\int\diff^4x\l[- \f{1}{48}F_{\m\n\rho\s}F^{\m\n\rho\s}
+ \f{q}{24}\eps^{\m\n\rho\s}\l(F_{\m\n\rho\s}-4\p_\m C_{\n\rho\s}\r) \r] 
+ S_{\rm b} \ ,
\ee
where 
\be
\label{eq:boundary_term}
S_{\rm b} = 
\f 1 6 \int\diff^4x\,\p_\m\l(q \eps^{\m\n\rho\s}C_{\n\rho\s}\r) \ ,
\ee
and $q = q(x)$ is a Lagrange multiplier with mass/scaling dimension two, \ie
under dilatations behaves as
\be
\label{eq:q_dilatations}
q(x) \to q'(x) = e^{2\s}q(e^\s x) \ ,
\ee
and enforces the Bianchi identity~(\ref{eq:field_strength}). 

Variation of~(\ref{eq:3form_first_order}) wrt $F$ yields
\be
\label{eq:F_eom}
F_{\m\n\rho\s} = q \eps_{\m\n\rho\s} \ ,
\ee
which upon plugging back into the action gives
\be
\label{eq:F_integrated_out}
S_{\rm 3-form} = 
\int\diff^4x \l[-\f{q^2}{2} 
+ \f 1 6 \eps^{\m\n\rho\s}C_{\n\rho\s}\p_\m q \r] \ ,
\ee
and we notice that the above is still scale invariant, owing
to~(\ref{eq:dilatations_CF},\ref{eq:q_dilatations}).

Nevertheless, a scale does appear, but only after the three-form equation of
motion
\be
\p_\m q = 0 \ ,
\ee
has been imposed, meaning that the spurion is forced to be a dimensionful
integration constant 
\be
q = q_0 \ ,
\ee
and dilatations are broken dynamically. 

The on-shell action is found to be 
\be
S_{\rm 3-form} = -\f{q_0^2}{2} \int \diff^4x  \ ,
\ee
consistent with the fact that the three-form theory does not propagate any
degrees of freedom.

\subsection{Three-form induced spontaneous scale symmetry breaking}
\label{sec:dilaton_basics}

The previous section has made clear that the three-form is fully compatible
with scale invariance and, most importantly, with its breaking, as it
dynamically provides a quantity of mass dimension two. 

This invites studying what happens when a real scalar with dimension
one---which we shall refer to as the~\emph{dilaton}---interacts with the gauge
sector in a dilatation-invariant manner. As we shall show, this coupling
generates a symmetry-breaking potential for the dilaton, lifting its flat
direction and inducing a nonzero vacuum expectation value.

At lowest order, the scale-invariant action involving the dilaton $\c$ and a
three-form reads (in the first order formulation)
\begin{align}
\label{eq:flat_dilaton_3form_combined}
S =  
\int\diff^4x \Bigg[ - \f 1 2 (\p_\m \c)^2 & - \f{\b}{4}\c^4
 - \f{1}{48}F_{\m\n\rho\s}F^{\m\n\rho\s} \nn \\
& - \f{1}{24}\sqrt{\f \lambda 2} \c^2 \eps^{\m\n\rho\s} F_{\m\n\rho\s} 
+ \f{q}{24}\eps^{\m\n\rho\s}\Bigl(F_{\m\n\rho\s}-4\p_\m C_{\n\rho\s}\Bigr)
\Bigg] + S_{\rm b} \ ,
\end{align}
where $\b$ and $\lambda>0$ are dimensionless; the boundary term $S_{\rm b}$
was given in~(\ref{eq:boundary_term}).\footnote{In the second-order
formulation, the boundary term in the presence of a coupling between a scalar
and a three-form reads~\cite{Giudice:2019iwl}
\be
\label{eq:dilaton_boundary}
S_{\rm b} = 
\f 1 6 \int \diff^4x\,\p_\m\l[\l(F^{\m\n\rho\s}
+ \sqrt{\f \lambda 2} \c^2 \eps^{\m\n\rho\s}\r)C_{\n\rho\s} \r] \ .
\ee} 

Owing to exact scale symmetry, the interaction between the dilaton and
three-form is~\emph{uniquely} fixed; $\c^2 \eps^{\m\n\rho\s} F_{\m\n\rho\s}$
is the only dimension-four mixing term simultaneously compatible with
dilatations and the gauge redundancy, as any other coupling would require
additional powers of $\c$ and/or derivatives, thus raising the operator
dimension beyond the lowest-order truncation. Note that such a coupling was
also employed in~\cite{Giudice:2019iwl} to enable the scanning, and eventually
the relaxation, of the scalar mass and cosmological constant. There, the
scalar in question was identified with the Standard Model Higgs field. More
broadly, these ideas are rooted in the earlier ``attractor vacua''
constructions of~\cite{Dvali:2003br,Dvali:2004tma}. The present work was
partly inspired by them. 

The equation of motion for $F$ is algebraic
\be
\label{eq:F_eom_dilaton}
F_{\m\n\rho\s} = 
\l(q - \sqrt{\f \lambda 2}\,\c^2\r) \eps_{\m\n\rho\s} \ ,
\ee
and substituting this into~(\ref{eq:flat_dilaton_3form_combined}) yields
\be
\label{eq:flat_F_integrated_out}
S = 
\int \diff^4x \l[ - \f 1 2 (\p_\m \c)^2
- \f{\lambda}{4}\l(\c^2 - \sqrt{\f{2}{\lambda}}\,q\r)^2 - \f{\b}{4}\c^4
+ \f 1 6 \eps^{\m\n\rho\s}C_{\n\rho\s}\p_\m q \r] \ .
\ee
As expected, the action at this stage is still invariant under dilatations.

The breakdown of scale symmetry in this setup proceeds in two steps. First,
exactly as in the free theory discussed in the previous section, the equations
of motion of the three-form supply the dimensionful integration constant
$q_0$, which we shall take to be positive. Second, this constant enters the
dilaton potential
\be
\label{eq:dilaton_potential}
V(\c) = \f{\lambda}{4}\l(\c^2-\sqrt{\f{2}{\lambda}}\,q_0\r)^2
+ \f{\b}{4}\c^4 \ ,
\ee
which develops a nontrivial minimum at
\be
\label{eq:extrema_flat}
\c_0^2 = \f{\sqrt{2\lambda}}{\lambda+\b} q_0 \ ,
\ee
provided $\lambda + \b > 0$. Thus, the flat direction is lifted and the
would-be Nambu-Goldstone mode acquires a nonzero mass
\be
\label{eq:mass_flat}
m_\c^2 = 2\sqrt{2\lambda}q_0 \ .
\ee

It is the coupling to the three-form that dynamically generates the scale
entering the potential for the dilaton and allows it to acquire a nonzero
vacuum expectation value without tuning away its quartic self-coupling. To
appreciate why this is nontrivial, note that had we considered the free
dilaton action in isolation
\be
\label{eq:free_dilaton_action}
S_{\rm dilaton} = 
\int \diff^4x \l[-\f 1 2 (\p_\m\c)^2 - \f{\b}{4} \c^4 \r] \ ,
\ee
a Poincaré-invariant ground state that spontaneously breaks scale symmetry
would require $\beta = 0$,\footnote{This is the scalar analogue of the
cosmological constant problem~\cite{Sundrum:2003yt}.} and this is the only
option if dilatations are an exact symmetry. In that case, physics is entirely
independent of the vacuum expectation value of $\c$, the ground state is
infinitely degenerate, and the dilaton remains massless.

Of course, relaxing the requirement of exact scale symmetry is always an
option, see \eg~\cite{Coradeschi:2013gda}. Then, it is the explicit breaking
of scale symmetry that lifts the flat directions, but to pinpoint how this
materializes is far from universal and requires additional assumptions
concerning how the breaking arises, how weak or strong it is, and what form
the scale-violating operators take.

On the other hand, the point of the present construction is that even if exact
invariance under dilatations is assumed, a scale can be generated dynamically
through the three-form sector, with the resulting potential for the dilaton
uniquely fixed and of quartic form.

This suggests that once the field is coupled nonminimally to gravity---as
required by scale invariance in curved spacetimes---the induced potential gets
exponentially flattened, raising the possibility that the dilaton could play
the role of the inflaton.

Note that what we have been discussing is closely related to unimodular
gravity~\cite{vanderBij:1981ym,Wilczek:1983as,Zee:1983jg,Buchmuller:1988wx,
Weinberg:1988cp,Unruh:1988in,Henneaux:1989zc,Alvarez:2005iy}---of course, such
a comparison becomes meaningful only once one passes to curved spacetimes, and
we shall return to it in the next section. In both cases, a dimensionful
integration constant is present and acts as the (sole) source of
scale-symmetry breaking. The difference lies in the form of the resulting
potential for the dilaton.

\section{Gravity}
\label{sec:gravity}

\subsection{Induced gravity}

For gravity to enter the picture in a manner maximally compatible with exact
scale symmetry,\footnote{For dynamical backgrounds, global scale
transformations can be written with coordinates fixed as
\bea
&&g_{\m\n}(x) \to g'_{\m\n}(x) = e^{-2\s}g_{\m\n}(x) \ ,
~~~\c(x) \to \c'(x) = e^\s \c(x) \ ,
~~~q(x) \to q'(x) = e^{2\s}q(x) \ ,\\
&&C_{\m\n\rho}(x) \to C'_{\m\n\rho}(x) = e^{-2\s}C_{\m\n\rho}(x) \ ,
~~~F_{\m\n\rho\s}(x) \to F'_{\m\n\rho\s}(x) = e^{-2\s}F_{\m\n\rho\s}(x) \ .
\eea} 
it is necessary to nonminimally couple the dilaton to the Ricci scalar
$R$.\footnote{We require that gravity propagates only the massless graviton.}
The corresponding action is a straightforward curved-space generalization
of~(\ref{eq:flat_dilaton_3form_combined}), and reads
\begin{align}
S =  
\int\diff^4x\sqrt{g} \Bigg[\f{\x \c^2}{2}R &- \f 1 2 (\p_\m \c)^2  
- \f{\b}{4}\c^4 - \f{1}{48}F_{\m\n\rho\s}F^{\m\n\rho\s} \nonumber \\
& - \f{1}{24}\sqrt{\f \lambda 2} \c^2 \mathcal E^{\m\n\rho\s} F_{\m\n\rho\s}
+ \f{q}{24}\mathcal E^{\m\n\rho\s}\Bigl(F_{\m\n\rho\s}
-4\nabla_\m C_{\n\rho\s}\Bigr)
\Bigg] 
+ S_{\rm b} \ ,
\end{align}
where $g = -{\rm det}(g_{\m\n})$, $\nabla_\m$ is the covariant derivative, and
$\mathcal E^{\m\n\rho\s} = \eps^{\m\n\rho\s}/\sqrt{g}$; the boundary term is
given by~(\ref{eq:boundary_term}) with the replacement $\eps^{\m\n\rho\s} \to
\mathcal E^{\m\n\rho\s}$. 

Integrating out $F$ and $C$, the theory reduces to a nonminimally coupled
scalar 
\be
S = 
\int \diff^4x\sqrt{g} 
\l[\f{\x \c^2}{2}R - \f 1 2 (\p_\m \c)^2 - V(\c) \r] \ ,
\ee
with the symmetry breaking quartic potential $V(\c)$ given
in~(\ref{eq:dilaton_potential}). Varying this wrt the metric and dilaton, we
obtain
\begin{align}
&\x\c^2 \l(R_{\m\n} - \f 1 2 g_{\m\n}R\r) = 
\p_\m\c \p_\n\c + \x\l(\nabla_\m \nabla_\n - g_{\m\n}\square\r)\c^2 
-g_{\m\n}\l(\f 1 2 (\p_\rho\c)^2+V(\c)\r) \ ,
\label{eq:Einstein_EOM} \\
&\square \c +\x \c R -V'(\c) = 0 \ ,
\label{eq:dilaton_EOM}
\end{align} 
respectively; $R_{\m\n}$ is the Ricci tensor, and $\square=g^{\m\n}\nabla_\m
\nabla_\n$ the covariant d'Alembertian. 

Focus on constant field configurations. Tracing~(\ref{eq:Einstein_EOM}), and
plugging into~(\ref{eq:dilaton_EOM}), we notice that in the presence of
gravity the ground state of the system is 
\be
\label{eq:dilaton_vev_curved}
\c_0^2 = \sqrt{\f 2 \lambda}q_0 \ , 
\ee
and does not depend on $\b$---as usual in scale-invariant settings, this
parameter only controls the cosmological constant
$\Lambda$~\cite{Shaposhnikov:2008xb,Garcia-Bellido:2011kqb,Coradeschi:2013gda},
as can be seen from~(\ref{eq:dilaton_vev_curved}) and~(\ref{eq:Einstein_EOM})
\be
\label{eq:CC}
\Lambda = \f{\b}{4\x}\c_0^2 \ . 
\ee
Minkowski background corresponds to $\b=0$, whereas de Sitter (anti de Sitter)
to $\b>0~(\b<0)$. 

The dynamics is better understood by getting rid of the mixing between the
dilaton and $R$, by Weyl-rescaling the metric as 
\be
g_{\m\n} \mapsto \f{\x \c^2}{M^2_{\rm Pl}}g_{\m\n} \ ,
\ee
where the reduced Planck mass $M_{\rm Pl}$ is related to the vacuum
expectation value~(\ref{eq:dilaton_vev_curved}) of the dilaton as
\be
M_{\rm Pl} = \sqrt{\x} \c_0 \ .
\ee

The resulting action is given by
\be
S = 
\int\diff^4x\sqrt{g}\l[ \f{M^2_{\rm Pl}}{2}R 
- \f{M^2_{\rm Pl}}{2}\l(6+\f{1}{\x}\r)\f{(\p_\m\c)^2}{\c^2} 
- \f{\lambda M^4_{\rm Pl}}{4\x^2}\l(1-\f{M^2_{\rm Pl}}{\x\c^2}\r)^2 
- \Lambda M^2_{\rm Pl}\r] \ ,
\ee
and after canonicalizing the kinetic term for the field by introducing 
\be
\tau=M_{\rm Pl}\sqrt{6+\f{1}{\x}}\log\f{\sqrt\x\c}{M_{\rm Pl}} \ , 
\ee
we end up with 
\be
\label{eq:Higgs_Starobinsky}
S = 
\int\diff^4x\sqrt{g}
\l[\f{M^2_{\rm Pl}}{2}R -\f 1 2 (\p_\m \tau)^2 
- \f{\lambda M^4_{\rm Pl}}{4\x^2}
\l(1-e^{-\f{2\tau}{M_{\rm Pl}\sqrt{6+1/\x}}}\r)^2 
- \Lambda M^2_{\rm Pl}\r] \ . 
\ee

As expected, the field has an exponentially flat potential that belongs to the
same universality class as the familiar Higgs~\cite{Bezrukov:2007ep} and
Starobinsky~\cite{Starobinsky:1980te} models. Clearly, a bare cosmological
constant $\Lambda \propto \b$, see~(\ref{eq:CC}), already leads to inflation
albeit without a graceful exit, irrespectively of any scalar sector. If,
however, inflation is to be driven by the dilaton, its potential energy must
dominate over the cosmological constant. In any event, the latter has to be
fine-tuned to be tiny---and this is not explained here. 

The inflationary observables $n_s$ and $r$, as well as the power spectrum
$\mathcal A_s$, to leading order in the number of efoldings $N \sim \mathcal
O(50-60)$ are
\be
n_s \simeq 1 - \f 2 N \ ,
~~~r \simeq \f{2}{N^2}\l(6+\f 1 \x\r) \ ,
~~~\mathcal A_s \simeq \f{\lambda}{6\pi^2\x^2 r} \ .
\ee
Requiring that the amplitude matches
observations~\cite{Planck:2018jri,BICEP:2021xfz} fixes 
\be
\f{\lambda}{\x^2\l(6+\f 1 \x \r)} \sim  10^{-10} \ ,
\ee
and in turn the dilaton mass is $m_\tau \sim \mathcal O (10^{-5})M_{\rm Pl}$.

Since $\lambda$ and $\xi$ are free parameters, the above condition determines
this particular combination of them. As can be immediately seen, in the limit
$\x\gg1$ the predictions are identical to Higgs/Starobinsky inflation.
Successful inflation in this context, however, is not restricted to this
regime, as it is also possible to take $\xi\sim \mathcal O(1)$, provided
$\lambda$ is correspondingly small.

At this point, it is worth mentioning the interesting
work~\cite{Csaki:2014bua}, where explicit breaking by nearly marginal
operators was shown to generate the exact same plateau structure; the present
construction achieves this without any explicit breaking, relying instead on
the three-form integration constant. A crucial ingredient in both cases is the
nonminimal coupling of the dilaton to gravity.

\subsection{Comparison with Unimodular gravity}
\label{sec:UG}

The above considerations should be contrasted with what happens in  
scale-invariant unimodular
gravity~\cite{Shaposhnikov:2008xi,Garcia-Bellido:2011kqb} or, more generally,
in theories with transverse
diffeomorphisms~\cite{Blas:2011ac,Karananas:2016grc}. There too dilatations
are broken by an integration constant and moreover the resulting dilatonic
potential is also uniquely fixed. In unimodular constructions, however, this
potential is of the runaway type. Accordingly, the dilaton is naturally
identified with a quintessence-like field responsible for the present-day
accelerated expansion of the Universe.\footnote{The interpretation of the
dilaton as a cosmon/quintessence field has a long history going back
to~\cite{Peccei:1987mm,Wetterich:1987fm,Wetterich:1994bg}, see
also~\cite{Wetterich:2019qzx,Wetterich:2022ncl} for recent reviews.} When
realistic setups are considered, such as the Higgs-dilaton
model~\cite{Shaposhnikov:2008xi,Garcia-Bellido:2011kqb}, the inflationary
epoch is governed by the Higgs field, while the dilaton becomes relevant at
late times and controls the Dark-Energy dynamics. As an added bonus, the late
and early dynamics are nontrivially related~\cite{Garcia-Bellido:2011kqb}.

At first sight, this may seem to contradict the standard lore that the
three-form and unimodular descriptions are dynamically equivalent. This
degeneracy is indeed present as long as one does not insist on exact scale
invariance: in the absence of the nonminimal coupling required by dilatations,
the two formulations are indistinguishable. The crucial point, however, is
that once the action is required to be exactly invariant under dilatations,
the admissible couplings are no longer the same, and the way the integration
constant enters the scalar sector becomes formulation-dependent. In this
sense, once exact dilatation invariance is imposed, the (naive) equivalence
between unimodular and three-form descriptions no longer survives at the level
of the effective scalar dynamics.

\subsection{A comment on $R^2$ gravity}

A dilaton-like scalar may also originate from scale-invariant gravity itself,
namely in the pure $R^2$ theory
\be
\label{eq:R2_bare}
S_{\rm R^2}= \f{1}{f^2}\int\diff^4x\sqrt{g}\,R^2 \ ,
\ee
with $f$ dimensionless.  Although this theory does not admit (Ricci-)flat
vacua perturbatively~\cite{Golovnev:2023zen,Karananas:2024hoh,Barker:2025gon},
it is classically equivalent to Einstein gravity with a strictly non-vanishing
cosmological constant, supplemented by a free massless
scalar~\cite{Alvarez-Gaume:2015rwa} (see also~\cite{Hell:2023mph}) associated
with the conformal mode of the metric. 

It is then logical to wonder what a purely gravitational counterpart of the
three-form lifting would yield. Consider the lowest-order scale-invariant
extension of the above 
\begin{align}
\label{eq:R2_3form_first_order}
S
=
\int\diff^4x\sqrt{g}
\Bigg[\f{1}{f^2} R^2- \f{1}{48}F_{\m\n\rho\s}F^{\m\n\rho\s}
&- \f{\sqrt{8\lambda}}{24}\,R\,\mc E^{\m\n\rho\s}F_{\m\n\rho\s} \nn\\
&+ \f{q}{24}\,\mc E^{\m\n\rho\s} 
\l(F_{\m\n\rho\s}-4\nabla_\m C_{\n\rho\s}\r)\Bigg] + S_{\rm b} \ ,
\end{align}
where, as previously, $S_{\rm b}$ can be found in~(\ref{eq:boundary_term})
with $\eps^{\m\n\rho\s} \to \mathcal E^{\m\n\rho\s}$. 

On the equations of motion for $F$ and $C$, the action reduces to
\be
\label{eq:R2_3form_fR}
S = \int\diff^4x\sqrt{g}
\l[ \f{M^2_0}{2}R + \a R^2-\f{M^4_0}{64\lambda} \r] \ ,
\ee
where
\be
\label{eq:R2_3form_gamma_def}
\a = \f{1}{f^2}-4\lambda \ ,
~~~M_0^2 = 2\sqrt{8\lambda}q_0 \ .
\ee
Notice that the three-form dynamically generated the Einstein-Hilbert term,
whose presence breaks scale symmetry.\footnote{Interestingly, this somewhat
resembles the situation in Weyl-invariant Einstein-Cartan gravity,
see~\cite{Karananas:2024xja,Karananas:2025xcv}, but in reverse. There, a
cross-term between the scalar and pseudoscalar curvatures induces a term
linear in the latter.}

Passing to Einstein frame and using $M_{\rm Pl}=M_0/\l(2\sqrt{\lambda}f\r)$,
yields a scalaron potential of Starobinsky type,
see~(\ref{eq:Higgs_Starobinsky}), but with an important caveat. In the $R^2$
case, the same parameter $f^2$ that controls the cosmological constant
$\Lambda = f^2 M^2_{\rm Pl}/16$ also fixes the height of the would-be
inflationary plateau. This difference from the dilaton considerations is
crucial. There, the inflationary scale is controlled by a parameter
combination independent of the vacuum energy. Here, on the other hand, if one
requires that $\Lambda$ reproduces the observed late-time vacuum energy, then
$f\lll 1$, and the exponential plateau gets suppressed far below what is
needed for viable inflation, unless another extreme tuning is imposed on
$\lambda$. One may of course adopt a somewhat~\emph{laissez-faire} attitude
towards the vacuum energy, and simply assume that additional sectors
contribute so that the observed late-time cosmological constant does not fix
$f$ by itself. In which case inflation is of course possible and with
predictions in excellent agreement with observations.

\section{Conclusions}
\label{sec:conclusions}

The three-form gauge field is well known to furnish a simple mechanism for the
dynamical generation of a dimensionful parameter. We showed that when a
dilaton is coupled to such a gauge sector in a way compatible with
dilatations, the corresponding integration constant feeds into the scalar
potential and lifts the otherwise flat direction. In this way, the scale that
governs the breaking of dilatations emerges dynamically from the three-form
sector, without the need for explicit scale-violating operators.

With gravity included, a nonminimal coupling of the dilaton to the Ricci
scalar,  as required by exact scale invariance, converts the induced quartic
potential into an exponentially flat one, placing the construction in the same
universality class as Higgs and Starobinsky inflation. Unlike constructions
based on explicit scale-symmetry breaking~\cite{Csaki:2014bua}, no small
parameter needs to be tuned to achieve flatness---the plateau is an exact
consequence of the three-form integration constant combined with the
nonminimal coupling.

What we discussed here also differs from the situation in scale-invariant
unimodular constructions, where the corresponding integration constant gives
rise instead to a runaway potential and a qualitatively different cosmological
role---that of dynamical Dark Energy---for the dilaton. In other words, once
exact scale invariance is imposed, the equivalence between the unimodular and
three-form descriptions no longer survives at the level of the effective
scalar dynamics.

\section*{Acknowledgments} 

It is a great pleasure to thank Gia Dvali, Alex Kehagias and Misha
Shaposhnikov for discussions and comments on the manuscript. 

\setlength\bibsep{5pt}
\bibliographystyle{utphys}
\bibliography{Refs.bib}

\end{document}